\documentclass[lettersize,journal]{IEEEtran} 
\usepackage{cite}
\usepackage[colorlinks]{hyperref}
\usepackage{mathrsfs}
\usepackage{amsthm}
\usepackage{mathtools}
 \usepackage{siunitx}
\usepackage{array}
\usepackage{eqparbox}
\usepackage{bm}

\usepackage{etoolbox} \makeatletter \patchcmd{\@makecaption} {\scshape} {} {} {} \makeatother
\usepackage[table,dvipsnames]{xcolor}
\usepackage{multicol,booktabs,tabularx}

\usepackage{cases}
\usepackage{algorithm}
\usepackage{paralist}
\usepackage{algpseudocode}

\usepackage{graphicx}
\usepackage{subfigure}
\usepackage{epstopdf}
\usepackage{epsfig}
\usepackage{amssymb}
\usepackage{array}
\usepackage{multirow}
\usepackage{makecell}
\usepackage{soul} 

\title{Effective outdoor pathloss prediction: A multi-layer segmentation approach with weighting map}

\begin{document}
\author{Yuan Gao, Tao Wen, Wenjing Xie, Jianbo Du, Yong Zeng, \textit{Fellow, IEEE}, Dusit Niyato, \textit{Fellow, IEEE}, Shugong Xu, \textit{Fellow, IEEE}
\thanks{This work is supported by Shanghai Natural Science Foundation under Grant 25ZR1402148. (Corresponding author: Shugong Xu)} 
\thanks{Yuan Gao, Tao Wen and Wenjing Xie are with the School of Communication and Information Engineering, Shanghai University, China, email: gaoyuansie@shu.edu.cn, wenty2023@shu.edu.cn and wenjingxie@shu.edu.cn.}
\thanks{Jianbo Du is with the School of Communication and Information Engineering, X'an University of Posts and Telecommunications, X'an, China, e-mail: dujianboo@163.com.}
\thanks{Yong Zeng is with the National Mobile Communications Research Laboratory, Southeast University, Nanjing, and also with the Purple Mountain Laboratories, Nanjing, China, e-mail: yong\_zeng@seu.edu.cn.}
\thanks{Dusit Niyato is with the College of Computing and Data Science, Nanyang Technological University, Singapore, e-mail: dniyato@ntu.edu.sg.}
\thanks{Shugong Xu is with Xi’an Jiaotong-Liverpool University, Suzhou, China, email: shugong.xu@xjtlu.edu.cn.}}

\maketitle

\begin{abstract}
Predicting path loss by considering the physical environment is crucial for effective wireless network planning. Traditional methods, such as ray tracing and model-based approaches, often face challenges due to high computational complexity and discrepancies between models and real-world environments. In contrast, deep learning has emerged as a promising alternative, offering accurate path loss predictions with reduced computational complexity. In our research, we introduce a ResNet-based model designed to enhance path loss prediction. We employ innovative techniques to capture key features of the environment by generating transmission (Tx) and reception (Rx) depth maps, as well as a distance map from the geographic data. Recognizing the significant attenuation caused by signal reflection and diffraction, particularly at high frequencies, we have developed a weighting map that emphasizes the areas adjacent to the direct path between Tx and Rx for path loss prediction. {Extensive simulations demonstrate that our model outperforms PPNet, RPNet, and Vision Transformer (ViT) by 1.2-3.0 dB using dataset of ITU challenge 2024 and ICASSP 2023. In addition, the floating point operations (FLOPs) of the proposed model is 60\% less than those of benchmarks.} Additionally, ablation studies confirm that the inclusion of the weighting map significantly enhances prediction performance. 
\end{abstract}


%
\IEEEpeerreviewmaketitle

\section{Introduction}

6G is envisioned to provide ultra-low latency, highly reliable and ubiquitous services for both communications and sensing \cite{gao2025stochastic,du2024secure,gao2024performance,xu2025enhanced}. Signal attenuation, i.e., path loss, significantly influences the design and performance of cellular networks \cite{lee2024scalable,zeng2021toward,zeng2024tutorial}. As wireless technologies advance, particularly with the emergence of 5G and beyond, precise predictions of path loss are vital for effective network planning and optimization. Without reliable forecasts, engineers face the risk of inadequate coverage or resource wastage, highlighting the necessity for robust methods to predict signal behavior in various outdoor environments \cite{zhou2024ample}.

Nonetheless, traditional path loss prediction methods present notable limitations. On one hand, model-based approaches, such as empirical formulas, often simplify real-world complexities to simple equations, missing the intricacies of environments like urban and suburban areas \cite{liu2025efficient}. This can lead to unreliable results when terrain or buildings alter signal paths. On the other hand, ray tracing (RT) methods are widely used for path loss simulation \cite{yang2024efficient}. {To obtain more accurate results, RT-based solutions require precise environmental modeling and material information to calculate the reflection and refraction ratios of electromagnetic waves. Although supported by modern GPU acceleration, computation time of a single simulation will also increase significantly as the number of sampled rays and effective reflections increases, which is not conducive to rapid inference.} These constraints render conventional methods less efficient for the rapid, large-scale demands of contemporary networks.  

Recently, artificial intelligence (AI) has become widely used to improve the performance of the cellular networks \cite{gao2025joint,jiang2025towards,jin2025linformer,gao2025multi,hu2024multi,gao2025ssnet}, and is also promising for improving path loss prediction, offering enhanced accuracy and efficiency \cite{hayashi2023deep,9977760}. Current AI-based approaches typically combine environmental information, primarily visual information, and extensive path loss datasets for path loss prediction, such as the channel knowledge map (CKM) \cite{zeng2021toward,zeng2024tutorial}. RadioUNet is a UNet-based deep learning model utilizing the geographical map and transmitter (Tx) location to predict the path loss \cite{levie2021radiounet}. \cite{qiu2022pseudo,hayashi2023deep} utilized convolutional neural networks (CNNs) to preprocess geographical data and model path loss.  \cite{lee2024scalable} leveraged building maps and Tx locations to calculate path loss to various Rx points, while \cite{sotiroudis2024deep} incorporated line-of-sight (LoS) and non-line-of-sight (NLoS) conditions to enhance accuracy. 
{However, these approaches rely heavily on data-driven methods and often overlook radio signal propagation characteristics, limiting their predictive accuracy. Specifically, the visual information of the scenario and buildings is generally used as the input of the AI-based models in a whole, and limited prior information regarding the signal propagation characteristics is exploited explicitly to enhance the performance of path loss prediction.}

{To address the above limitations, we propose a deep-learning based model that integrates the channel propagation-related information explicitly extracted from the 3D environmental map for effective path loss prediction. Specifically, from the 3D environmental map, we extracted the map of Tx depth, map of Rx depth and distance map, which influence the propagation of wireless signals dramatically. In addition, considering the fact that signal attenuation via reflections and diffraction are significant, especially for high frequencies, we propose a weighting map to give larger weight to the area near the direct path between the Tx and Rx. It is worth noting that the above features used in our model provide the essential prior information for path loss prediction, which is consistent with the explanatory variables employed in classical empirical models. In particular, our learning-based framework can be viewed as a data-driven extension that adapts the mapping functions of traditional propagation models, making them more aligned with realistic channel characteristics. This ensures that our feature design and learning strategy are not arbitrary, but rather grounded in well-established radio propagation principles. The performance comparison using the dataset in ITU AI/ML in 5G Challenge\footnote{https://challenge.aiforgood.itu.int/match/matchitem/96\label{challenge}} shows that the proposed model outperforms the existing benchmark by over 1.2 dB with 60 \% less floating point operations (FLOPs). The proposed model further outperforms by over 2.5 dB using the dataset in ICASSP 2023 \cite{bakirtzis2025}.}

\section{Problem Formulation}
In an outdoor environment with multiple objects, the received signal is modeled as the superposition of $N$ multipath components (MPCs) as:

\begin{align}
    h(t,\tau,\varOmega,\varPsi)
    = \sum_{m=1}^{N} \alpha_m \,
      \delta(\tau-\tau_m)\,
      \delta(\varOmega-\varOmega_m)\,
      \delta(\varPsi-\varPsi_m),
\end{align}where $\alpha_m$, $\tau_m$,  $\varOmega_m$ and $\varPsi_m$ are the complex amplitude, propagation delay, and angles of departure and arrival, respectively. Accordingly, the path loss $\mathrm{PL}$ (in dB) is calculated as:
\begin{align}
    \mathrm{PL} = -10\log_{10}(\sum_{m=1}^{N} |\alpha_m|^2). 
\end{align}

As $\alpha_m$ depends on distance attenuation, shadowing, and interactions with surrounding objects (e.g., reflections), path loss is a deterministic function of the Tx/Rx positions and the surrounding environment as:
\begin{equation}\label{eq:mapping}
    \mathrm{PL} = f({x},\, {y},\, \mathcal{E}),
\end{equation}
where $x$, $y$ and $\mathcal{E}$ denote the Tx, Rx positions, and environment, respectively. For traditional path loss prediction methods, such as empirical channel model, $\mathcal{E}$ generally denotes the type of propagation environment and $f()$ is an analytical function. For AI-based path loss prediction, $\mathcal{E}$ generally contains much richer information of the environment, such as the visual information, and $f(*)$ is the learnable mapping between the Tx, Rx positions, environment information and the value of path loss. $f(*)$ is learned by training towards the following mean squared error minimization:

\begin{small}
\begin{equation}
\min_{f(*)}\text{MSE} = {\frac{1}{N} \sum_{i=1}^{N} \left( \mathrm{PL}_i - \widehat{\mathrm{PL}_i} \right)^2},
\label{eq:rmse}
\end{equation}\end{small}where \(\hat{PL}_i=f(x,\, y_i,\, \mathcal{E})\) is the predicted path loss (dB) of the $i$-th Rx, \({PL}_i\) is the ground truth path loss (dB) of the $i$-th Rx, \(N\) is the total number of training samples.
\begin{figure}[htbp]
    \centering
    \includegraphics[width=0.7\columnwidth]{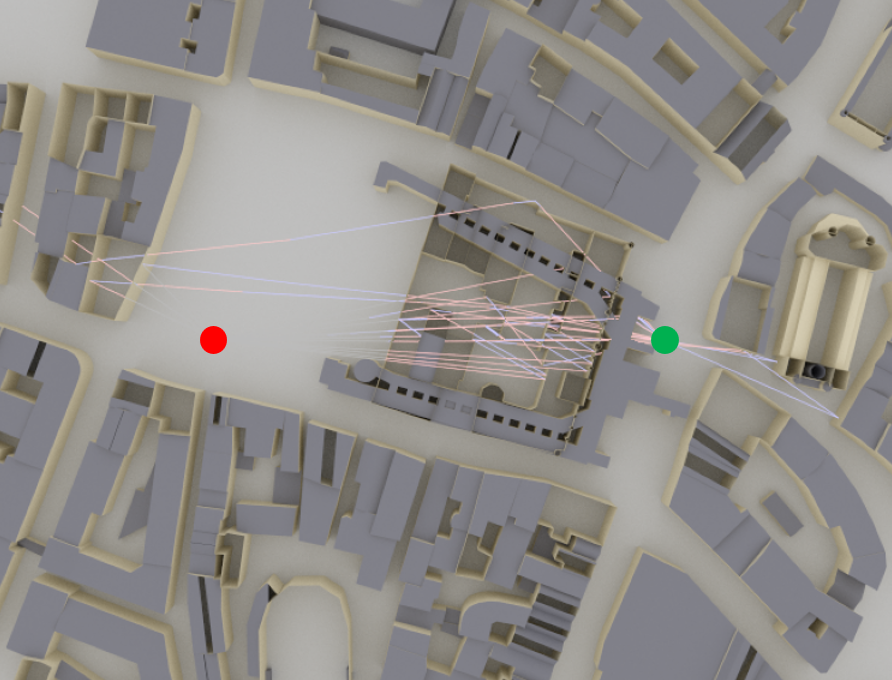}     
    \caption{Illustration of the path distribution (red lines) between Tx (red point) and Rx (green point) using RT model, where only the rays with relatively high power at the Rx are kept.}
    \label{fig:RT}
\end{figure}

\begin{figure*}
\centering
\includegraphics[width=1\linewidth]{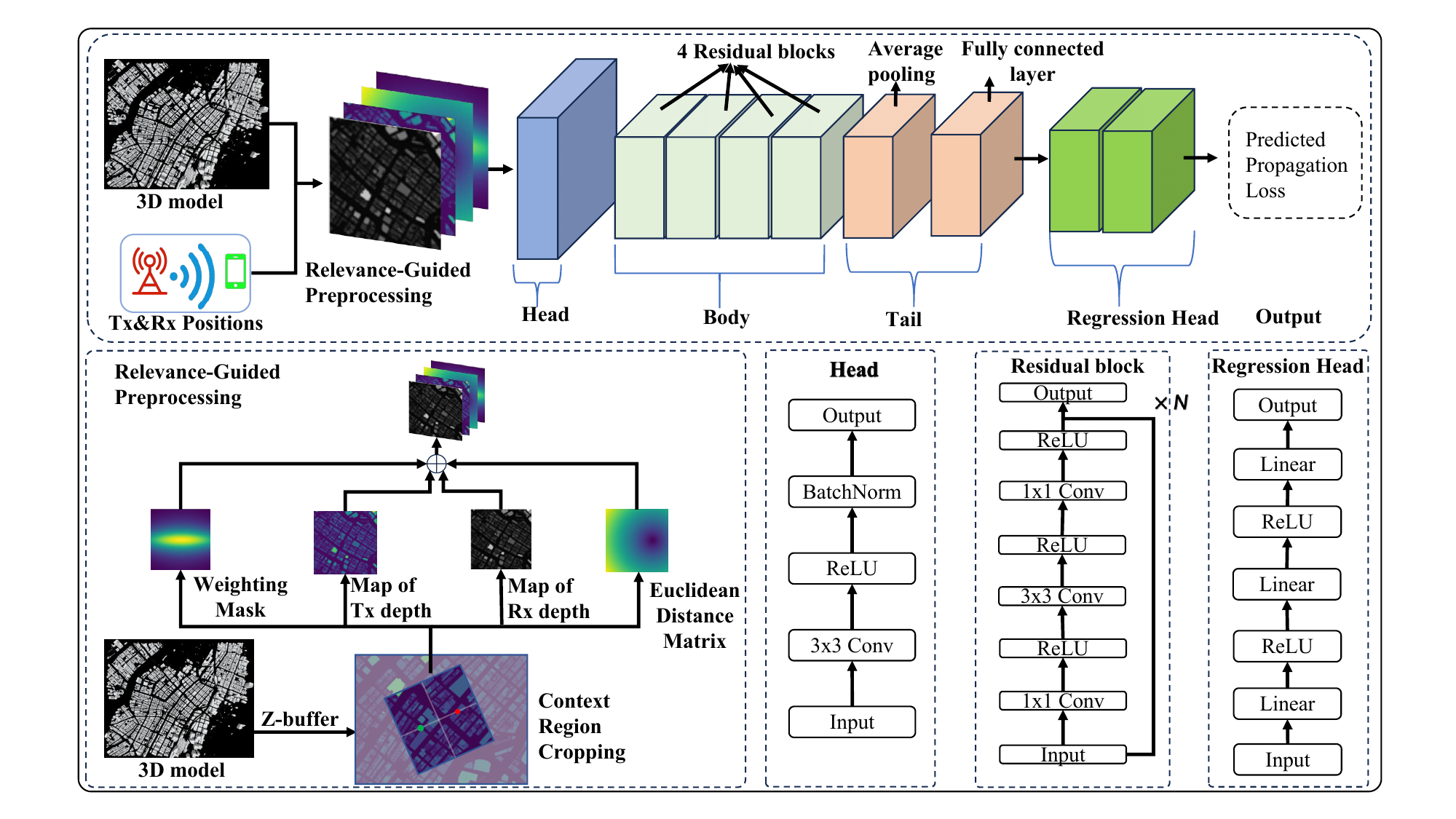} 
\vspace{-20px}
\caption{{Overview of the path loss prediction architecture.
In the data preprocessing stage, the environmental features are decomposed into the Tx-depth map, Rx-depth map, and distance map, followed by a \textbf{Context Regional Cropping} procedure. This procedure extracts a square region that encloses the Tx–Rx area together with its surrounding context, which serves as the reference region for subsequent feature extraction. In addition, a weighting map is introduced to assign higher importance to areas located near the direct propagation path between the Tx and Rx.
The proposed model consists of a ResNet-based backbone for multi-level feature extraction and a regression head that estimates the path loss. }}
\label{fig:model_architecture}
\end{figure*}

\section{Proposed Model}

\subsection{Relevance-Guided Preprocessing}
To effectively capture the characteristics influencing path loss, including Rx position, Tx position, relative distance, and environmental factors, the complete 3D map is decomposed into multiple layers, which serve as inputs to the proposed model.

\textbf{Context Region Cropping:}
{As illustrated in Fig.~\ref{fig:model_architecture}, to mitigate the influence of irrelevant variables and the relative positions of Tx and Rx in the input feature map, as well as to reduce learning complexity, we crop a square region covering the Tx–Rx path and its surrounding area as in Fig.~\ref{fig:model_architecture}. The Tx and Rx are positioned at the first and third quartile points along the horizontal axis. Subsequent feature maps are derived from this region by resizing to a fixed resolution of $H\times W$ as specified in Table~\ref{tab:training_config}, while absolute distance information is separately provided by the Distance Map described later.}

\textbf{Map of Rx Depth:} 
The depth of the Rx relative to surrounding buildings is calculated by subtracting the Rx height from the building height, i.e., {$Rx_\text{depth} = building_\text{height} - Rx_\text{height}$}. A positive $Rx_\text{depth}$ indicates that the Rx is below the building height, suggesting a high probability of NLoS conditions. Conversely, a negative $Rx_\text{depth} $ suggests the Rx is above the building, implying likely LoS. Apart from LoS/NLoS identification, the map of Rx depth provides a richer, more granular understanding of the environment around Rx. Utilizing the distribution of Rx depth allows us to infer the obstruction caused by buildings, thereby aiding in more accurate path loss prediction.

\textbf{Map of Tx Depth:} 
Similarly, the depth of the Tx relative to buildings is computed as: $Tx_\text{depth} = building_\text{height} - Tx_\text{height}$. Similarly,for LoS/NLoS identification, map of Rx depth provides a richer, more granular understanding of the environment around Tx. Incorporating the distribution of Tx depth enhances our understanding of environmental obstructions affecting signal propagation.

\textbf{Distance Map:} The distance map is generated based on the Euclidean distance between the Tx (located at the origin) and each Rx position. This map is used to learn the nonlinear relationship between path loss and the Tx-Rx distance, enabling the model to account for the impact of distance on signal attenuation.

\textbf{Weighting Map:} {From the experimental perspective, RT simulations (Fig.~\ref{fig:RT}) reveal that rays with relatively high received power are clustered in the vicinity of the direct path, while contributions from other areas decay rapidly after reflection and diffraction. This observation inspires our Gaussian-inverse-square weighting mask is a principled mechanism grounded in both radio propagation theory and empirical evidence.} The path loss between the Tx and Rx is the aggregation of the signal propagating in multiple propagation paths, and the attenuation of the reflective and diffractive paths is significant, especially for high frequencies. This means that the environment features in the region near the direct path between the Tx and Rx dominate in the contribution of path loss. To model this influence, we introduce a weighting mask generated using a Gaussian distribution \cite{qiao2021gaussian} along the x-axis and an inverse-square distribution along the y-axis, centered on the region of interest (RoI) at $(H/2,W/2)$ within the input map (size $H \times W$). The weighting map assigns higher weights to areas close to the direct path, emphasizing their importance, and lower weights to areas farther away, reducing their influence. This mask emphasizes the areas most relevant to accurate path loss prediction. 

{Mathematically, the two-dimensional distribution of the weighting mask is defined as  
\begin{small}
\begin{equation}
W(x, y) = \mathcal{G}(x; \sigma) \cdot \mathcal{I}(y; \kappa),
\end{equation}
\end{small}where the Gaussian component along the \(x\)-axis, 
\begin{small}
\begin{equation}
\mathcal{G}(x; \sigma) = \exp\left(-\frac{x^2}{2\sigma^2}\right),
\end{equation}
\end{small}captures the fact that contributions near the direct Tx--Rx path dominate, while the inverse-square component along the \(y\)-axis,  
\begin{small}
\begin{equation}
\mathcal{I}(y; \kappa) = \frac{1}{1 + {y^2}/{\kappa}},
\end{equation}
\end{small}attenuates the influence of regions further away from the LoS corridor. With hyper-parameters \(\sigma\) and \(\kappa\) in Table.~\ref{tab:training_config}, the mask emphasizes the physically meaningful region where effective multipath contributions concentrate.}

\subsection{Deep Learning-Based Model}

To enable effective path loss prediction, we developed a deep learning framework illustrated in Fig. \ref{fig:model_architecture}. The core idea is to learn a nonlinear mapping function expressed in Eq. (\ref{eq:mapping}).

To effectively extract these environmental features, we adopt a ResNet-based backbone \cite{wu2019wider}, which balances training complexity and feature extraction capacity. Its multi-level feature extraction and atrous convolution design make it particularly well-suited for path loss prediction, outperforming conventional CNN architectures. The model consists of the following key components:
\begin{itemize}
    \item \textbf{Head}: A 3×3 convolutional layer that performs initial feature extraction, followed by a batch normalization layer to stabilize and accelerate training, and a ReLU activation function to introduce non-linearity.
    \item \textbf{Body}: Composed of four residual block groups, each comprising multiple bottleneck residual blocks. Skip connection within each block mitigates the vanishing gradient problem, enabling training of deeper networks. Bottleneck residual blocks utilize three convolutional layers to optimize parameter efficiency \cite{prince2023understanding}. The initial layer uses a 1×1 kernel to decrease the number of channels, followed by a standard 3×3 kernel, and finally, another 1×1 kernel to restore the channel count to its original size.
    \item \textbf{Tail}: An adaptive average pooling layer functions to reduce the size of the input feature map by taking the average value within a local region. This process helps to decrease computational complexity, reduce the number of parameters, and improve the model's robustness to translational variations. The output is then fed into a fully connected layer that maps the features to the environmental representation.  

\end{itemize}

Following the backbone, a regression head maps these extracted features directly to the predicted path loss. This head includes three fully connected layers: the first applies a linear transformation to the input features, transforming the features into an intermediate representation with ReLU activation; the second refines this representation further, preparing them for the final output. The second ReLU activations introduce non-linearity, allowing for better modeling of complex patterns in the data. The last linear layer produces the continuous path loss prediction.

\begin{figure}[htbp]
    \centering
    \includegraphics[width=0.8\columnwidth]{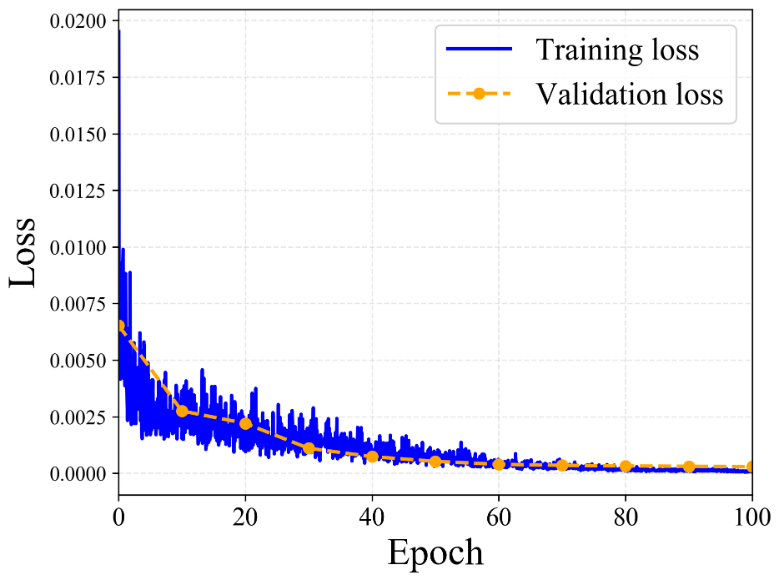}     
    \caption{{Convergence analysis of the proposed model. 
    Both training and validation loss curves indicate fast and stable convergence.}}
    \label{fig:convergence}
\end{figure}

\subsection{Model Training}
The hyper-parameters and training configurations for the proposed model are summarized in Table \ref{tab:training_config}. Training was performed on an NVIDIA GeForce RTX 3090 GPU. To stabilize the learning process, input data, including depth maps and coordinate information, was normalized by dividing each value by 250. During inference, the model output predictions were de-normalized by multiplying by 250, converting the estimated path loss back to the original unit of simulation (dB). This ensures that the root mean square error (RMSE) accurately reflects the true path loss prediction error. {We illustrate both the training and the validation loss in Fig.~\ref{fig:convergence}, which shows that our model exhibits fast convergence within the first 60 epochs and maintains stable performance thereafter, demonstrating the training stability of our proposed model.}

\begin{table}[htbp]
\centering
\caption{Training configurations and hyper-parameters of the proposed model}
\label{tab:training_config}
\begin{tabular}{lcc}
\toprule

{Dataset}& \multicolumn{2}{c}{ITU challenge 2024\textsuperscript{\ref{challenge}}, ICASSP 2023 \cite{bakirtzis2025}}\\
Split for training (test) set& \multicolumn{2}{c}{75\% (25\%) of dataset}\\
Learning rate (LR) & \multicolumn{2}{c}{$10^{-3}$}\\
LR gamma, step size & \multicolumn{2}{c}{0.8, 8}\\
Region H & \multicolumn{2}{c}{80}\\
Region W & \multicolumn{2}{c}{80}\\
Weight map \(\sigma\) &   \multicolumn{2}{c}{40}  \\
Weight map \(\kappa\) &   \multicolumn{2}{c}{150} \\
Batch size & \multicolumn{2}{c}{64}\\
Optimizer & \multicolumn{2}{c}{Adam} \\
Number of epochs & \multicolumn{2}{c}{100} \\
CV 1& Training: Tx 1, 2, 3; Testing: Tx 4\\
CV 2& Training: Tx 1, 2, 4; Testing: Tx 3\\
CV 3& Training: Tx 1, 3, 4; Testing: Tx 2\\
CV 4& Training: Tx 2, 3, 4; Testing: Tx 1\\
RT 1 & Samples: \(1 \times 10^6\); depth: 4\\
RT 2 & Samples: \(1 \times 10^6\); depth: 10\\
\bottomrule
\end{tabular}
\end{table}

\begin{table*}
\centering
\caption{{Performance comparison between the proposed model and benchmark models using the dataset in ITU challenge 2024\textsuperscript{\ref{challenge}}.}}
\label{Perforamnce_comparison}
\begin{tabular}{llcccccccc} 
\toprule
Area&Model&RMSE(dB)&MAE(dB)&MAPE($\%$)&R\textsuperscript{2}&FLOPs(G)& Time(s)&Param(M)&Memory(MB)\\
\hline
\multirow[c]{5}{*}{\makecell[c]{1Km \\ $\times$ \\ 2Km}}&PPNet \cite{qiu2022pseudo}   & 13.46           & 12.43           & 10.75         & 0.48         & 10.10         &0.00123          &56.91          &140.32\\
&RPNet \cite{9977760}        & 14.65           & 13.65           & 11.92         & 0.45         & 8.40          &0.00101          &43.75          &132.51\\
&ViT \cite{han2022survey}-6   & 13.94           & 12.19           & 12.67         & 0.46         & 18.10         &0.0029           &97.40           &431.03\\
&ViT \cite{han2022survey}-12  & 13.77           & 12.05           & 10.23         & 0.43         & 36.10         &0.0045           &186.80          &812.24\\
&Ours      & \textbf{12.19}  &\textbf{10.12}   & \textbf{9.14} & \textbf{0.58} &\textbf{4.10}  &\textbf{0.00086} &\textbf{23.75} &\textbf{105.20}\\
\hline
\multirow[c]{3}{*}{\makecell[c]{200m \\ $\times$ \\ 200m}}
 &RT 1& 7.17  & 5.70  & 5.92  & 0.85 & - &4.53& - & - \\
&RT 2& 6.36  & 5.61  & 5.74  & 0.89 & - &8.21& - & - \\
&Ours& \textbf{4.28} & \textbf{3.92}  & \textbf{3.71}  &  \textbf{0.92} & 4.10&\textbf{0.00086} &23.75 &105.20\\
\bottomrule
\end{tabular}
\end{table*}

\begin{table}[h]
\centering
\caption{{The performance evaluation of the proposed model in ICASSP 2023 \cite{bakirtzis2025}}.}
\begin{tabular}{lcccc}
\toprule
Model&RMSE(dB)&MAE(dB)&MAPE($\%$)&R\textsuperscript{2}\\
\hline
PPNet \cite{qiu2022pseudo} &8.09&5.66&5.42  &0.83 \\
RPNet \cite{9977760}       &9.56&7.20&6.01 &0.80 \\
ViT \cite{han2022survey}-12&8.72&6.34&5.83 & 0.76\\
Ours&\textbf{5.59}&\textbf{4.16}&\textbf{3.97}&\textbf{0.88}\\
\bottomrule
\end{tabular}
\label{tab:ICASSP}
\end{table}



\section{Simulation Results Analysis}
\subsection{Simulation Settings}
We compare the proposed model with state-of-the-art benchmark models using two datasets. The first one is the large-scale outdoor path loss dataset used in ITU AI/ML in 5G challenge\textsuperscript{\ref{challenge}}, which was generated using the 3D buildings in the area of Chuo and Chiyoda wards, Tokyo. In this scenario, four Tx are placed at strategic locations, and the path loss values at various Rx coordinates (specified by latitude and longitude) are computed for each Tx. The path loss was obtained using a RT model that accounts for up to four reflections and one diffraction, capturing the complex propagation phenomena in urban environments. The volume of data per Tx at each frequency is approximately 36,000 records, resulting in a total of around 432,000 data points across all Txs and frequencies. The second dataset is used in 2023 ICASSP for the first pathloss radio map prediction challenge 

As illustrated in Table \ref{tab:training_config}, these are the average results of our method on the ITU dataset across all cross-validation (CV). For each CV, the data of three Tx are used as the training set, and the pathloss data of the remaining one Tx are used as the test set. The dataset partition strategy is summarized in Table \ref{tab:training_config}, ensuring that each Tx point is included exactly once. The path loss prediction error is quantified using:
\begin{small}
\begin{equation}
\text{RMSE} = \sqrt{\frac{1}{N} \sum_{i=1}^{N} \left( PL_i - \widehat{PL_i} \right)^2},
\label{eq:rmse}
\end{equation}\end{small}where \(\hat{PL}_i\) is the predicted path loss (dB) of the $i$-th Rx, \({PL}_i\) is the ground truth path loss (dB) of the $i$-th Rx obtained using RT simulation, \(N\) is the total number of testing samples. In addition, we comprehensively evaluated the performance of our method using Mean Absolute Error (MAE), Mean Absolute Percentage Error (MAPE), and Coefficient of Determination \(R\textsuperscript{2}\).

\subsection{Performance Comparison}


To validate the effectiveness of the proposed model, we compare its performance against two baselines: the classical path loss prediction models PPNet \cite{qiu2022pseudo}, RPNet \cite{9977760} and the state-of-the-art image processing model, Vision Transformer (ViT) \cite{han2022survey}. PPNet \cite{qiu2022pseudo} is based on a fully convolutional encoder-decoder architecture designed to perform pixel-level transformations of input images to target outputs via supervised learning. It is widely regarded as a standard baseline for path loss prediction. {RPNet \cite{9977760} is also based on convolution and residual modules, but its input feature maps only consider the environmental maps around Tx  and Rx separately, lacking the overall spatial structure information.} ViT, built on the Transformer architecture, demonstrates remarkable image processing capabilities by leveraging the self-attention mechanism for long-range feature extraction. {To ensure fair comparison, we retrained all competing models on the same dataset. Specifically, we followed the respective data preprocessing methods of PPNet and RPNet when retraining these two models, while the ViT network was trained using the same data preprocessing approach as our proposed method. Data splitting was kept completely consistent across all models.} 

As illustrated in Table \ref{Perforamnce_comparison}, the proposed model with Gaussian weight masking consistently achieves the lowest RMSE. On average, its RMSE is over 1.2 dB lower than that of PPNet, while requiring 60\% fewer FLOPs. The high computational demand of PPNet is due to its fully convolutional structure, which substantially increases as the input size and model depth grow. Specifically, PPNet utilizes an up-sampling module before every two consecutive convolutional layers, significantly expanding its FLOPs for each layer. Additionally, PPNet's depth is considerably greater than that of our model. {In terms of the overall network architecture selection, RPNet \cite{9977760} is most similar to our base model. However, its feature extraction method is more limited to the local maps centered on Tx and Rx, lacking attention to the entire propagation path. Its average metrics are 16.8\% lower than ours, which also demonstrates the effectiveness of our data feature preprocessing. }Although ViT excels in image processing, it underperforms in path loss prediction without tailored adjustments. Notably, the scaling law that benefits ViT in image tasks appears less effective in this context, likely because ViT is not specifically designed for path loss estimation. {In addition, we compare the proposed model with RT simulations in Table \ref{Perforamnce_comparison}. It should be noted that, since only terrain maps are available while material information is absent, the simulations rely on default material parameters. This inherent limitation of the simulation setup introduces discrepancies from the ground truth. In Table \ref{Perforamnce_comparison}, since RT-based methods exhibit a non-negligible probability of convergence failure at long distances, our evaluation is restricted to valid path loss samples within a 200 m \(\times\) 200 m area around each transmitter. Under generic settings without material details or parameter tuning, our method achieves performance comparable to RT simulations, demonstrating both accuracy and robustness. 

{Further, to validate the cross-scenario generalization capability of our method, we conducted comparisons with related works on the ICASSP Challenge 2023 dataset \cite{bakirtzis2025} in Table \ref{tab:ICASSP}. The proposed model outperforms the PPNet \cite{qiu2022pseudo}, RPNet\cite{9977760} and ViT \cite{han2022survey} dramatically. On this dataset, the proposed method significantly outperforms the PPNet on both RMSE and MAPE metrics. Specifically, the RMSE is reduced to 5.59 dB, 2.5 dB improvement over PPNet’s 8.09 dB, while the MAPE decreases to 3.97\(\%\), achieving a 26.8\(\%\) reduction compared to PPNet’s 5.42\(\%\). These quantitative results demonstrate the superior error control capability of the proposed approach.}



\begin{table}[h]
\centering
\caption{Ablation experiments: the effectiveness of the distance map and weighting mask (units: dB)}
\label{Ablation_experiments}
\begin{tabular}{ccccccc}
\hline
Frequency & Model          & CV 1   & CV2    & CV 3   & CV4    & Avg    \\ \hline
\multirow{3}{*}{800 MHz}  & w/o dist       & 44.56  & 34.21  & 32.12  & 42.12  & 38.25  \\
                          & w/o mask       & 12.88  & 12.87  & 13.10  & 12.71  & 12.90  \\
                          & proposed model & \textbf{11.77} & \textbf{12.31} & \textbf{12.80} & \textbf{12.52} & \textbf{12.35} \\
\multirow{3}{*}{7 GHz}    & w/o dist       & 43.13  & 52.32  & 48.21  & 56.21  & 49.96  \\
                          & w/o mask    & 13.02 & 13.09 & 13.46 & 13.07 & 13.16  \\
                          & proposed model & \textbf{11.72} & \textbf{12.05} & \textbf{12.66} & \textbf{12.30} & \textbf{12.19} \\
\multirow{3}{*}{28 GHz}   & w/o dist       & 51.87  & 56.89  & 64.21  & 64.12  & 59.27  \\
                          & w/o mask    & 13.73 & 13.69 & 13.80 & 13.26 & 13.63  \\
                          & proposed model & \textbf{12.10} & \textbf{12.28} & \textbf{13.02} & \textbf{12.48} & \textbf{12.47} \\ \hline
\end{tabular}
\end{table}

\subsection{Ablation Experiments}
In this section, we investigate the necessity of the distance map in path loss prediction and the effectiveness of the Gaussian weighting map in improving the results. As shown in Table \ref{Ablation_experiments}, {once the absolute distance scale provided by the distance map is removed, relying solely on relative height information from the environment fails to achieve accurate convergence in path loss estimation. This phenomenon further underscores the physical interpretability and rationality of our feature design, which is firmly grounded in geometric propagation laws.} Table \ref{Ablation_experiments} further illustrates that the inclusion of the weighting mask results in notable performance improvements quantitatively, especially at higher frequencies. Specifically, the model achieves improvements of 0.55 dB, 0.97 dB, and 1.16 dB at 800 MHz, 7 GHz, and 28 GHz, respectively. This performance gain can be attributed to the design of the weighting mask, which emphasizes the region along the direct path between the Tx and Rx, while assigning lower weights to areas farther from this line. The radio propagation is heavily related to its frequency, for higher frequency, the propagation loss is dominated by line-of-sight (LoS) propagation, while radio reflection and diffraction attenuate so heavily that they can be weakened.

\section{Conclusions}
In this work, we have proposed a deep learning-based approach that effectively captures environmental features by segmenting the 3D scenario map into the Tx depth map, Rx depth map, and distance map. Additionally, recognizing that reflections and diffraction significantly influence signal attenuation, particularly at higher frequencies, we have introduced a weighting map that emphasizes the area along the direct path between the Tx and Rx. Extensive simulations demonstrate that our model outperforms PPNet, PRNet, and ViT by 1.2-3.0 dB using dataset of ITU challenge 2024 and ICASSP 2023. In addition, FLOPs of the proposed model are 60\% less than those of benchmarks. Additionally, ablation studies confirm that the inclusion of the weighting map significantly enhances prediction performance. The ablation experiments have further confirmed that the weighting map substantially enhances prediction accuracy at higher frequencies, aligning with the design rationale. 

{In the future, we will work on the generalization, which is a key performance indicator for the applicability of the proposed model. Although the training and testing datasets are simulated with respect to the Txs of different physical environments, they share a certain level of similarities. We plan to collect more datasets via simulation and measurement from different scenarios, including noise, mobility and building structures, and focus on improving the generalization capability of the proposed model, for example we will explore how to refine the weighting map design across diverse scenarios and frequencies. }

\ifCLASSOPTIONcaptionsoff
  \newpage
\fi                                   
\bibliographystyle{IEEEtran}
\bibliography{bibfile}     

\end{document}